\documentclass[aps,prl,reprint,floatfix,citeautoscript,superscriptaddress]{revtex4-1}
\usepackage[utf8]{inputenc}
\usepackage{geometry}
\usepackage{setspace}
\usepackage{times}
\usepackage[T1]{fontenc}
\usepackage{amsmath}
\usepackage{hyperref} %参考文献跳转，此宏包会自动载入graphicx
\hypersetup{colorlinks=true,linkcolor=blue,anchorcolor=blue,citecolor=blue,urlcolor=black}
\usepackage{indentfirst}
\usepackage{graphicx}
\usepackage{booktabs}
\usepackage{makecell}
\usepackage{gensymb}
\usepackage{verbatim}
\usepackage{booktabs}
\usepackage{threeparttable}
\usepackage{multirow}
\usepackage{footnote}   % WW,
\usepackage{graphicx}   % WW,
\usepackage{appendix}   % WW,附录
\usepackage{placeins}   % Keep figures/tables inside their logical sections.
\usepackage[strings]{underscore} %处理文件名中的下划线

\usepackage{amssymb}
\usepackage{mathptmx}
\usepackage{bm}

\newcommand{\IHEP}{\affiliation{Institute~of~High~Energy~Physics, Chinese Academy of Sciences, Beijing, China}}
\newcommand{\UCAS}{\affiliation{University of Chinese Academy of Sciences, Beijing, China}}

\begin{document}

\title{A unified quenching model in Geant 4 simulation for $\alpha$, proton and electron particles in liquid scintillator detectors}

\author{Shubing Liu}\IHEP\UCAS
\author{Mingyuan Wang}\IHEP\UCAS
\author{Zeyuan Yu}\altaffiliation{Corresponding author: yuzy@ihep.ac.cn}\IHEP

% \date{\today}

\begin{abstract}

The liquid scintillator (LS) detectors are widely used in reactor neutrino experiments. To precisely measure the oscillated neutrino spectrum, it is crucial to understand the LS energy response, especially the quenching effect.
Numerous bench-top measurements have been conducted on quenching effects for $\alpha$ particles, protons, and electrons. These results have typically been described by Birks' law, but with different Birks' coefficients required for different particle species.
In this study, we find that if more secondary electrons are allowed to be generated in Geant4 simulations, the bench-top results for all particles can be well fitted using a single Birks' coefficient of about $0.013~{\rm g\,cm^{-2}\,MeV^{-1}}$.
The underlying reason is that a large fraction of the primary energy is deposited through the generation of $\delta$-electrons with energies below 4~keV, which should be tracked separately due to their different quenching behavior compared to primary particles.
This study provides a unified framework for comparing Birks' coefficients from different bench-top measurements and helps experiments like JUNO and SNO+ better tune their simulations for precision energy measurements.

\end{abstract}
\maketitle
% \linenumbers

\section{Introduction}

Organic liquid scintillator (LS) detectors have played a central role in neutrino experiments. Linear
alkylbenzene (LAB) has become a particularly important scintillator solvent
because of its high light yield, good optical transparency, chemical stability,
and compatibility with large detector volumes.  LAB-based or LAB-like
scintillators have been used or developed for Daya Bay \cite{Dayabay_LS}, JUNO \cite{JUNO_LS}, SNO+ \cite{SNO+_dissertation}, RENO \cite{RENO}, Double Chooz \cite{Double_Chooz_dissertation}.

For these detectors, the conversion from deposited energy of particles in LS to visible energy is
not perfectly linear.  A charged particle loses energy through ionization and
excitation, but only part of this deposited energy appears as prompt
scintillation light.  The reduction of scintillation efficiency at large
ionization density is usually called ionization quenching.  This effect is one
of the main ingredients of the non-linear energy response.
Its accurate description is important for liquid-scintillator experiments in
general, because ionization quenching affects the visible-energy scale, energy
resolution, particle identification, and detector-response modeling.

The canonical description of ionization quenching is Birks'
law.  In its one-parameter form, the scintillation light produced per unit track
length is written as
\begin{equation}
    \frac{dL}{dx}
    =
    S\frac{dE/dx}{1+kB\,dE/dx},
    \label{eq:introduction-birks}
\end{equation}
where $kB$ is Birks' coefficient \cite{Birks_Law}.  A commonly used
generalization introduced by Chou adds a second-order term
\cite{Chou_Law},
\begin{equation}
    \frac{dL}{dx}
    =
    S\frac{dE/dx}{1+kB\,dE/dx+C(dE/dx)^2},
    \label{eq:introduction-chou}
\end{equation}
where $L$ is the scintillation light output in photons, $x$ is the track length
in cm, $E$ is the deposited energy in MeV, and $S$ is the unquenched light
yield in photons/MeV.  In this standard form, $kB$ and $C$ have units of
$\mathrm{cm\,MeV^{-1}}$ and $\mathrm{cm^2\,MeV^{-2}}$, respectively.  In this work, we
report density-normalized form coefficients by using $(dE/dx)/\rho$ in the
quenching terms, giving units of $\mathrm{g\,cm^{-2}\,MeV^{-1}}$ for $kB$.  The standard values can be obtained from density-normalized values by dividing $kB$ by $\rho$, where $\rho$ is the density of the scintillator.
Besides the model of Chou, other models with more additional terms have also been developed based on canonical Birks' law
\cite{different_quenching_models}.

Experimentally, Birks' coefficient is often extracted by integrating
Eq.~\eqref{eq:introduction-birks} or Eq.~\eqref{eq:introduction-chou} with tabulated stopping powers, such as ESTAR \cite{ESTAR} and SRIM \cite{SRIM}.
 These
stopping-power-integral fits usually give different effective $kB$ values for
electrons, protons, and $\alpha$ particles; for the strongly quenched
$\alpha$ response, the second-order Chou term is often introduced to describe
the full energy range .
Previous bench-top measurements and detector-response studies have pointed out
that such coefficients are not directly transferable to Geant4 simulations,
because secondary-particle production and tracking are not represented in the
same way \cite{JUNO_energy_resolution, Dayabay_e-_kb_numerical_integration, Double_Chooz_e-_article, SNO+_proton_alpha_article, yangMeasurementProtonQuenching2019}.
Specifically, a stopping-power integral assigns the energy loss to the primary
particle trajectory, whereas Geant4 generates and tracks above-threshold
secondary electrons separately and applies quenching along their tracks.  The
secondary electrons have different local stopping powers from the primary
particle, so the energy loss is not represented equivalently in the two
approaches.

In this work, we extend this simulation-definition
issue from electron response to a combined comparison of electron, proton, and
$\alpha$ quenching data,  which is usually used for energy response studying in neutrino experiments.  By lowering the secondary-electron production
threshold and applying Birks' law step by step to both primary and secondary
tracks, we find that the apparent particle dependence of $kB$ is strongly
reduced.  Within this Geant4 step-level definition, the three particle
responses can be described with canonical Birks' law which includes only one quenching parameter, and the second-order Chou term or other additional terms is not needed for describing low-energy and high-energy $\alpha $ particle quenching effect simultaneously.

\section{Review of bench-top measurements}

The SNO+ experiment provides a systematic set of proton and $\alpha$ measurements
in LAB-based scintillators.  The proton light response was measured at the PTB
neutron beam for four LAB cocktails with different PPO and bis-MSB
concentrations \cite{SNO+_e-_proton_article}.  The recoil-proton spectra were
interpreted with the NRESP simulation, and the extracted light-output points
were fitted with Birks' law, or its Chou extension, using SRIM stopping
powers.  The fitted proton values were
$kB=(0.0094-0.0098)~{\rm cm\,MeV^{-1}}$, while the quadratic coefficient was
consistent with zero.  The same neutron-beam program was extended to
$\alpha$ particles from $^{12}$C$(n,\alpha){}^{9}$Be reactions in the same
LAB-based samples \cite{SNO+_dissertation}.  For the LAB sample with 2~g/l PPO and
15~mg/l bis-MSB, the one-parameter $\alpha$ fit gave
$kB=(0.0076\pm0.0003)~{\rm cm\,MeV^{-1}}$, compared with about
$0.0096~{\rm cm\,MeV^{-1}}$ for protons in the same scintillator.  An
independent SNO+ $\alpha$ study using neutron-induced $\alpha$ production, a
$^{147}$Sm-loaded sample, and radon-chain $\alpha$ peaks found
$kB=(0.0066-0.0076)~{\rm cm\,MeV^{-1}}$ \cite{SNO+_proton_alpha_article}.  In two
LAB samples where both proton and $\alpha$ responses were available, the
proton and $\alpha$ one-parameter Birks coefficients differed by $6.4\sigma$
and $5.8\sigma$, respectively.  This result showed that the traditional
stopping-power-integral coefficient is not particle universal even within the
same scintillator.

For JUNO liquid scintillator, the proton response was measured with a 14~MeV
D--T neutron generator by tagging recoil protons through the scattered neutron
angle from $20^\circ$ to $75^\circ$ \cite{yangMeasurementProtonQuenching2019}.  The visible scale
was anchored with the 0.321~MeV Compton electron from a $^{22}$Na source.  The
proton data were fitted by integrating Birks' law with a 1~$\mu$m numerical
step and SRIM stopping powers, with pull terms for the electron energy scale
and the neutron-beam direction.  The best fit was
$kB=(7.50\pm0.19)\times10^{-3}~{\rm g\,cm^{-2}\,MeV^{-1}}$ and
$C=(2.03\pm1.0)\times10^{-6}~{\rm g^2\,cm^{-4}\,MeV^{-2}}$.  The authors noted that the proton data were only
weakly sensitive to the $C$ term and that the proton $kB$ differed from the
electron value then used in the JUNO response model.

The JUNO energy-resolution study later used low-energy electron quenching data
below 0.2~MeV to determine the Birks coefficient directly within the JUNO
Geant4 simulation \cite{JUNO_energy_resolution}.  The electron inputs consisted of
the IHEP/Zhang Compton-scattering points and a TUM Compton-scattering dataset.
For each electron energy, Geant4 was used to simulate the energy deposition in
JUNO liquid scintillator, and the visible energy was calculated by summing
Birks' law over all steps of primary and secondary tracks in the event.  With
the JUNO production cuts kept fixed, the combined fit gave
$kB=(12.1\pm0.3)\times10^{-3}~{\rm g\,cm^{-2}\,MeV^{-1}}$.  The paper
emphasized that $kB$ values obtained from ESTAR or SRIM integrals cannot be
directly inserted into detector simulations, because those integrals do not
describe the production and independent tracking of secondary particles.

In the Daya Bay detector-calibration analysis, the scintillation nonlinearity
was modeled with Birks' law and constrained together with Cherenkov light and
electronics nonlinearity, giving
$kB=(15.2\pm2.7)\times10^{-3}~{\rm g\,cm^{-2}\,MeV^{-1}}$
\cite{Dayabay_e-_kb_numerical_integration}.  Although this was a detector-level calibration rather
than a pure bench-top extraction, it explicitly showed that the Geant4
secondary production threshold changes the simulated electron quenching curve:
lowering the electron cut produces more tracked $\delta$ electrons and a
different effective $kB$.

Double Chooz measured both the low-energy electron response and the
$\alpha$ response of its scintillator samples.  The electron article measured
20--140~keV Compton electrons for several Double Chooz scintillators
\cite{Double_Chooz_e-_article}.  A numerical integration using the Berger-Seltzer
electron stopping power gave
$kB=(0.0158\pm0.0006)~{\rm cm\,MeV^{-1}}$ for the Double Chooz
$\nu$-Target scintillator, while the other measured scintillator formulations
gave different best-fit values.  The same electron data were then fitted with Geant4.  Changing the
step-function parameters, energy-loss fluctuations, and the secondary
production cut changed the effective $\nu$-Target value from
$0.0107$ to $0.0287~{\rm cm\,MeV^{-1}}$, more than a factor of two.  The
Double Chooz thesis measured $\alpha$ quenching using internal $^{210}$Po,
$^{222}$Rn, $^{218}$Po, and $^{214}$Po sources and tuned the Monte Carlo with
combined electron and $\alpha$ data \cite{Double_Chooz_dissertation}.  The combined
$\alpha$ tuning gave $kB=(0.0098\pm0.0003)~{\rm cm\,MeV^{-1}}$ for the
Target scintillator and $kB=(0.0134\pm0.0005)~{\rm cm\,MeV^{-1}}$ for the
Gamma Catcher sample with 2~g/l PPO.  The thesis noted that, for
$\alpha$ particles, the light-yield shape is much less sensitive to $kB$
because of the very large stopping power, so the electron light scale is
needed to constrain the fit.

These measurements show a consistent pattern.  When Birks' law is fitted as a
continuous stopping-power integral, LAB-based proton data, LAB-based
$\alpha$ data, and low-energy electron data prefer different effective
coefficients, and the difference can be statistically significant even within
the same scintillator.  At the same time, electron-response studies and the
Double Chooz Geant4 tuning demonstrate that the fitted coefficient is not only
a material parameter, but also depends on how secondary particles and energy
deposition steps are represented.  This motivates a combined Geant4
step-level analysis in which the electron, proton, and $\alpha$ measurements
are compared under one common simulation definition.

\section{Simulation configuration and investigation}

The simulations were performed with Geant4 of version 11.4.2 \cite{Geant4}. For
the quenching calculation, the active region is treated as a cylindrical
LAB-like scintillator volume with a radius of 1982~mm and a height of
3964~mm, large enough to contain the charged-particle tracks studied in this
work.  All energy-deposition steps in the active scintillator are treated with
the same effective LAB-based liquid scintillator.  This simplification is
adequate for the present charged-particle study because optical-photon
transport is not simulated and the response quantity is the Birks-corrected
deposited energy.

The simulated material density is $0.859~{\rm g\,cm^{-3}}$.  Its elemental
mass fractions are 87.794\% C, 12.140\% H, 0.034\% O, 0.027\% N, and
0.005\% S.
This effective material was chosen to represent the common elemental
composition of LAB-based liquid scintillators.  The relevant Daya Bay
LS, JUNO LS, and SNO+ LAB1 samples all use LAB as the solvent and differ mainly
in the small concentrations of PPO and bis-MSB.
The published recipes and the corresponding material quantities used in this
comparison are summarized in Table~\ref{tab:ls-composition-comparison}.    Since PPO and bis-MSB are present only at the
g/L or mg/L level, their effect on the bulk C/H composition is small.
Therefore the same effective LAB-based material can be used as a controlled
simulation reference when comparing with electron, proton, and $\alpha$
quenching measurements from these LAB-based experiments.  Differences in
scintillator recipe are treated as an external material systematic rather than
as a change of the simulation definition.

\begin{table}[!htb]
    \centering
    \footnotesize
    \renewcommand{\arraystretch}{1.12}
    \begin{tabular}{@{}lll@{}}
        \toprule
        Scintillator & Density & Elements \\
        \midrule
        \makecell[tl]{Simulated LS\\
        \footnotesize (effective LAB-based)} &
        $\rho=0.859~\mathrm{g\,cm^{-3}}$ &
        \makecell[tl]{C 87.79\%, H 12.14\%\\
        O 0.034\%, N 0.027\%\\
        S 0.005\%} \\
        \addlinespace[0.65em]
        \makecell[tl]{Daya Bay LS  \cite{Dayabay_LS} \\
        \footnotesize (LAB + 3~g/L PPO\\
        \footnotesize + 15~mg/L bis-MSB)} &
        $\rho=0.859~\mathrm{g\,cm^{-3}}$ &
        \makecell[tl]{C 87.78\%, H 12.17\%\\
        O 0.025\%, N 0.022\%} \\
        \addlinespace[0.65em]
        \makecell[tl]{JUNO LS \cite{JUNO_LS, JUNO_energy_resolution}\\
        \footnotesize (LAB + 2.5~g/L PPO\\
        \footnotesize + 3~mg/L bis-MSB )} &
        $\rho=0.859~\mathrm{g\,cm^{-3}}$  &
        \makecell[tl]{C 87.79\%, H 12.17\%\\
        O 0.021\%, N 0.018\%} \\
        \addlinespace[0.65em]
        \makecell[tl]{SNO+ LAB1 sample  \cite{SNO+_dissertation} \\
        \footnotesize (99.766~wt\% LAB\\
        \footnotesize + 0.232~wt\% PPO\\
        \footnotesize + 0.002~wt\% bis-MSB)} &
        $\rho=0.863~\mathrm{g\,cm^{-3}}$ &
        \makecell[tl]{C 87.79\%, H 12.18\%\\
        O 0.017\%, N 0.015\%} \\
        \bottomrule
    \end{tabular}
    \caption{
    LAB-based liquid-scintillator recipes and material properties
    relevant to the present comparison.  The simulated active scintillator uses
    one effective elemental material.  Experimental elemental fractions are
    calculated from the published formulations listed in the table.  Here
    wt\% denotes mass fraction by weight.
    }
    \label{tab:ls-composition-comparison}
\end{table}

Primary particles are generated with a single-particle source.  The particle
type and kinetic energy are set separately for each sample.  For the response
samples discussed in this paper, the source position is fixed at the detector
center used in the geometry and the initial direction is fixed along the
negative $z$ axis.  The particle species simulated for the quenching study are
electrons, protons, and $\alpha$ particles.

The physics list is a custom modular list, summarized in
Table~\ref{tab:physics-list}. Low-energy electromagnetic interactions of electrons with kinetic energy below 0.1 MeV are modeled using the Livermore package.  The Livermore electron and photon models
are quoted as valid over 250~eV--100~GeV, usable down to 100~eV with reduced
accuracy, and in principle extendable to about 10~eV \cite{Livermore}.

\begin{table}[!htb]
    \centering
    \footnotesize
    \renewcommand{\arraystretch}{1.12}
    \begin{tabular}{@{}ll@{}}
        \toprule
        Component & Geant4 physics constructor/model \\
        \midrule
        Electromagnetic constructor &
        \makecell[tl]{G4LivermoreIonisationModel\\
        G4MollerBhabhaModel\\} \\
        \addlinespace[0.35em]
        Decay &
        \makecell[tl]{G4DecayPhysics\\G4RadioactiveDecayPhysics} \\
        \addlinespace[0.35em]
        Hadronic elastic &
        G4HadronElasticPhysicsHP \\
        \addlinespace[0.35em]
        Hadronic inelastic &
        G4HadronPhysicsQGSP\_BERT\_HP \\
        \addlinespace[0.35em]
        Stopping particles &
        G4StoppingPhysics \\
        \addlinespace[0.35em]
        Ion interactions &
        G4IonPhysicsPHP \\
        \bottomrule
    \end{tabular}
    \caption{
    Geant4 physics constructors registered in the custom modular physics list
    and the electron-ionization models selected by the electromagnetic
    constructors.
    }
    \label{tab:physics-list}
\end{table}

The secondary-production threshold is controlled by the electron and positron
range cuts.  The  parameter used in the production samples sets the
electron and positron cuts to 1, 10, or 100~$\mu$m, while the gamma range cut is
kept at its default value of 1~mm.  No other range cut is applied.
These Geant4 cuts are range cuts and are converted internally into
material-dependent kinetic-energy thresholds for secondary production.  Thus,
for ionization, the electron/positron cut controls whether a low-energy
secondary electron is generated as an explicit track or is included in the
restricted continuous energy loss of the parent particle.  Lowering the
electron/positron cut therefore resolves more low-energy delta electrons as
separate tracks, while increasing the cut merges a larger fraction of their
energy into the parent-particle step.

Optical physics is not registered because this work studies the scintillation
quenching energy, not optical-photon production, transport, or detection.  For
each Geant4 step in the active scintillator, the simulation takes the local
deposited energy $\Delta E$ and step length $\Delta x$, evaluates
$\delta=(\Delta E/\Delta x)/\rho$, and accumulates the quenched deposit as
\begin{equation}
\Delta E_{\rm vis} =
\frac{\Delta E}{1+kB\,\delta}.
\label{eq:simulation-step-birks}
\end{equation}
The visible energy is the sum of $\Delta E_{\rm vis}$ over all primary and
secondary steps in the active scintillator.

Our simulations were performed with the settings described above.  We first
examine the step-level information, using 7.68~MeV $\alpha$ particles, 6~MeV
protons, and 0.2~MeV electrons as representative samples for the investigation.

\begin{figure}[!htb]
    \centering
    \begin{minipage}{0.49\linewidth}
        \centering
        \includegraphics[width=\linewidth]{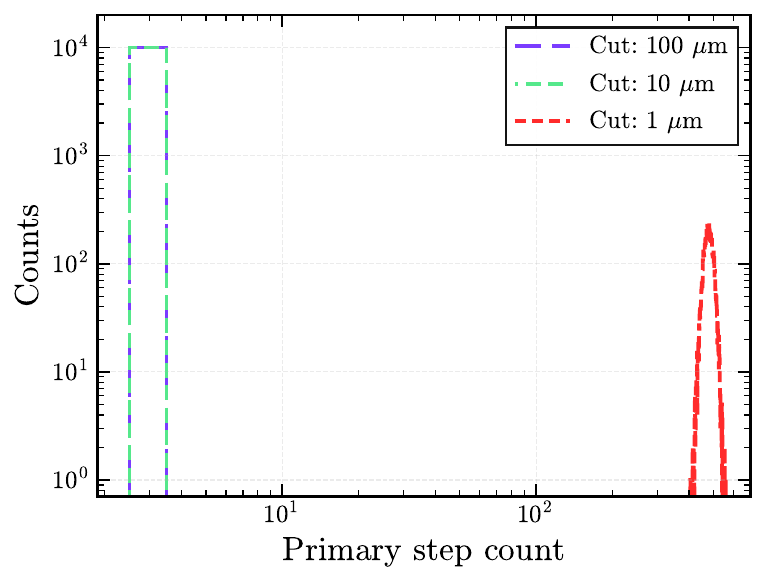}
    \end{minipage}
    \hfill
    \begin{minipage}{0.49\linewidth}
        \centering
        \includegraphics[width=\linewidth]{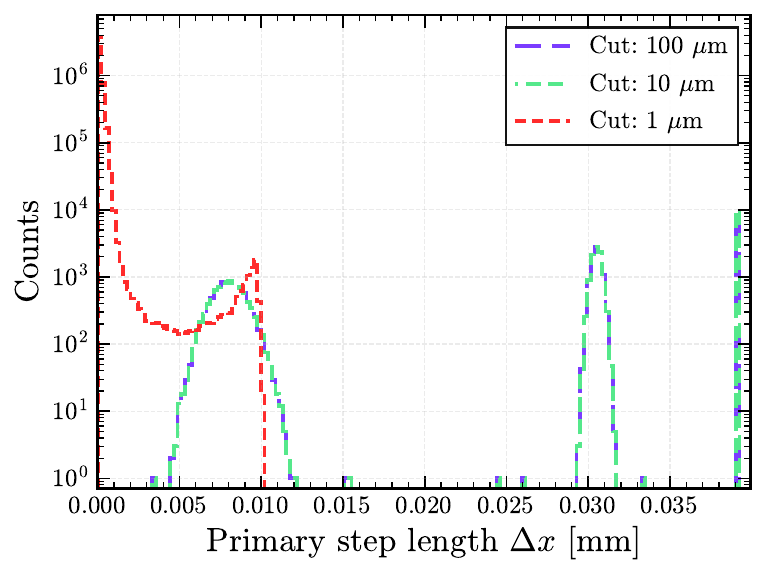}
    \end{minipage}
    \caption{
    Primary-track diagnostics for 7.68~MeV $\alpha$ particles.  The left panel shows the event-by-event number of
    primary steps, and the right panel shows the primary-step length
    distribution.
    }
    \label{fig:primary-step-alpha}
\end{figure}

\begin{figure}[!htb]
    \centering
    \begin{minipage}{0.49\linewidth}
        \centering
        \includegraphics[width=\linewidth]{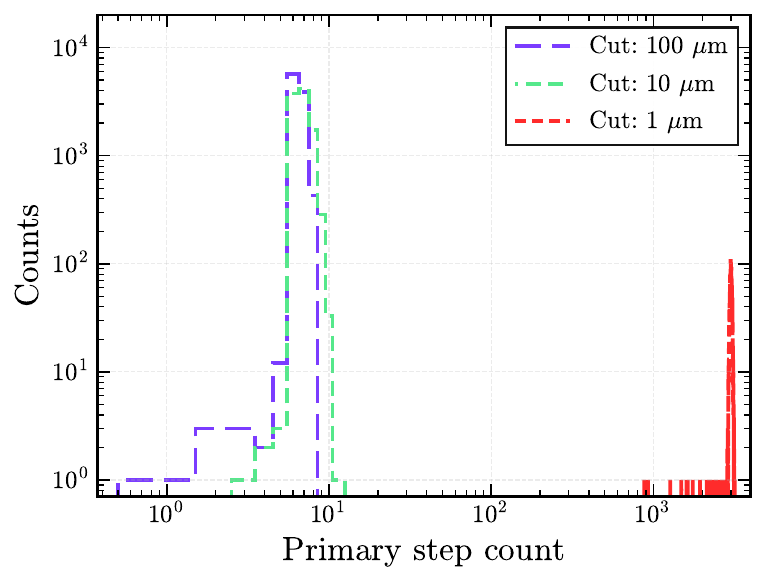}
    \end{minipage}
    \hfill
    \begin{minipage}{0.49\linewidth}
        \centering
        \includegraphics[width=\linewidth]{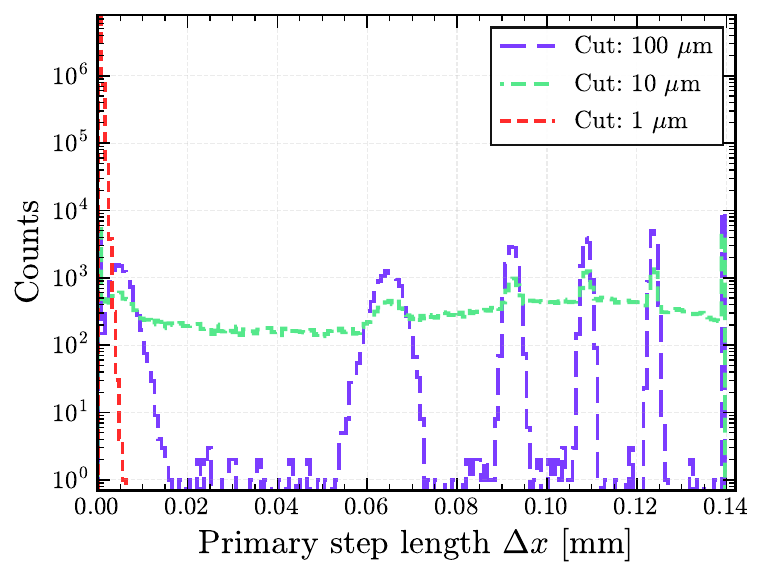}
    \end{minipage}
    \caption{
    Primary-track diagnostics for 6~MeV protons.  The left panel shows the event-by-event number of
    primary steps, and the right panel shows the primary-step length
    distribution.
    }
    \label{fig:primary-step-proton}
\end{figure}

\begin{figure}[!htb]
    \centering
    \begin{minipage}{0.49\linewidth}
        \centering
        \includegraphics[width=\linewidth]{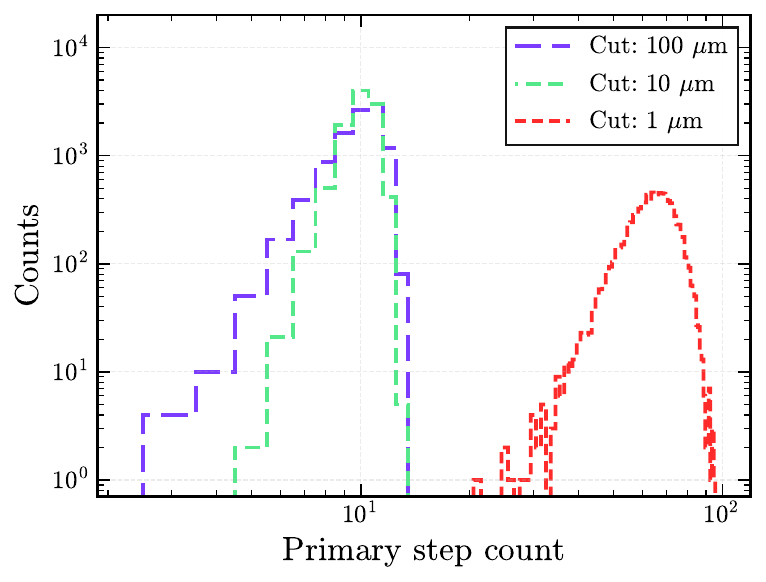}
    \end{minipage}
    \hfill
    \begin{minipage}{0.49\linewidth}
        \centering
        \includegraphics[width=\linewidth]{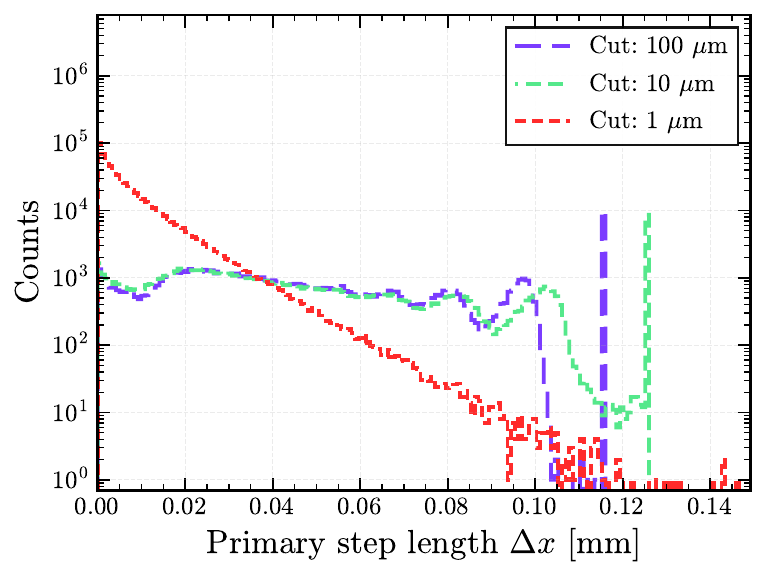}
    \end{minipage}
    \caption{
    Primary-track diagnostics for 0.2~MeV electrons. The left panel shows the event-by-event number of
    primary steps, and the right panel shows the primary-step length
    distribution.
    }
    \label{fig:primary-step-electron}
\end{figure}

We begin with the primary-track structure under different electron/positron
production cuts.  Figure~\ref{fig:primary-step-alpha}, \ref{fig:primary-step-proton}, and \ref{fig:primary-step-electron}  shows the number of
primary steps and the primary-step length distribution.    With
production cuts of 100 and 10~$\mu$m, no secondary
electrons are explicitly produced by the primary $\alpha$ track in this sample,
and the primary track is represented by only three or four long steps in most
events.  
When the cut is lowered to 1~$\mu$m, more low-energy secondary electrons are
created as independent tracks, so part of the energy loss is transferred from
the primary track to secondary-electron tracks.

\begin{figure*}[!htb]
    \centering
    \includegraphics[width=0.95\textwidth]{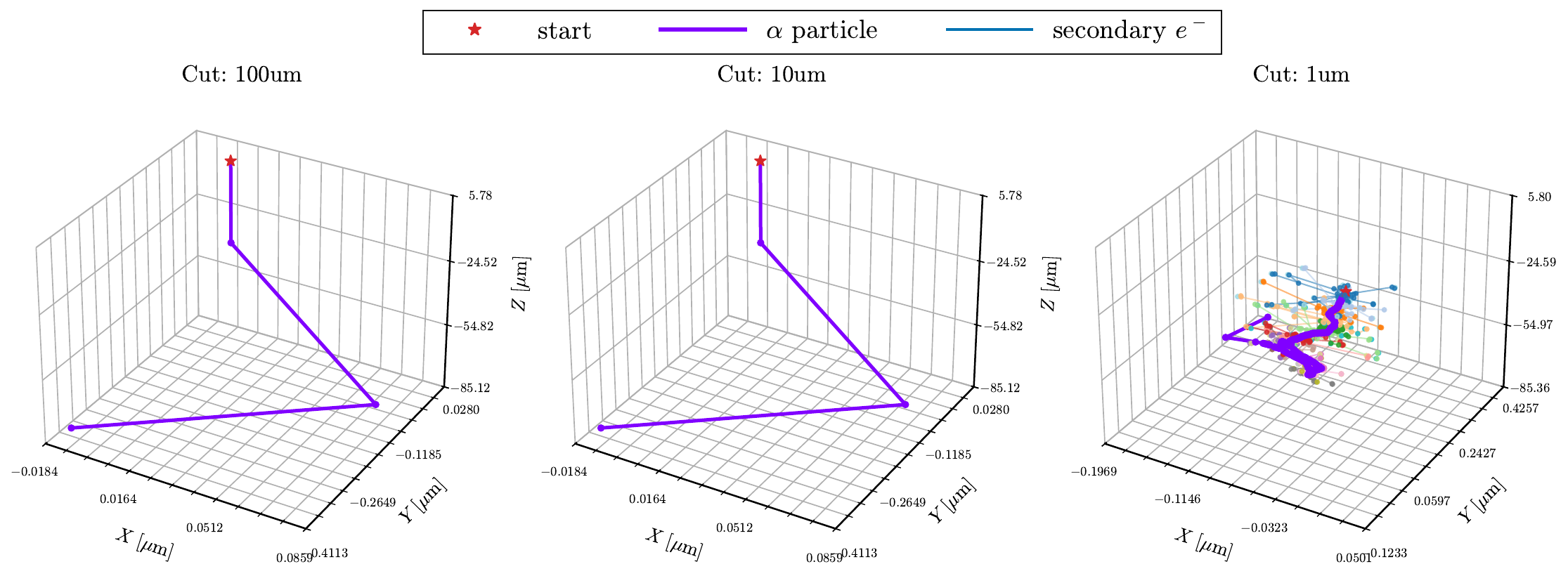}
    \caption{
    Representative 7.68~MeV $\alpha$-particle event displayed in three
    dimensions for electron/positron production cuts of 100, 10, and
    1~$\mu$m.  The start point, primary $\alpha$ track, and secondary-electron
    tracks are indicated in the legend.
    }
    \label{fig:track3d-alpha}
\end{figure*}

The same behavior is visible at the event level.  Figure~\ref{fig:track3d-alpha}
shows a representative 7.68~MeV $\alpha$ event for the three production cuts.
For the 100 and 10~$\mu$m cuts, the $\alpha$ track is a short sequence of long
primary steps, whereas the 1~$\mu$m cut produces explicit secondary-electron
tracks around the primary ionization column.

\begin{figure}[!htb]
    \centering
    \begin{minipage}{0.49\linewidth}
        \centering
        \includegraphics[width=\linewidth]{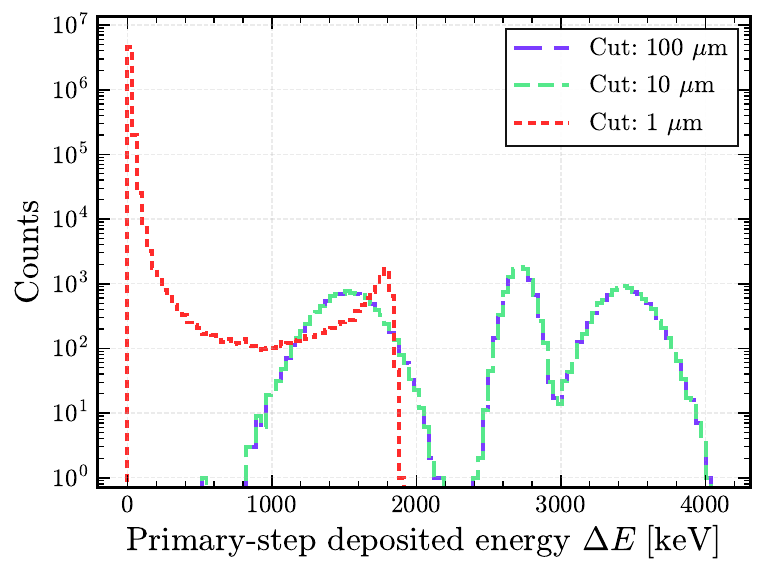}
    \end{minipage}
    \hfill
    \begin{minipage}{0.49\linewidth}
        \centering
        \includegraphics[width=\linewidth]{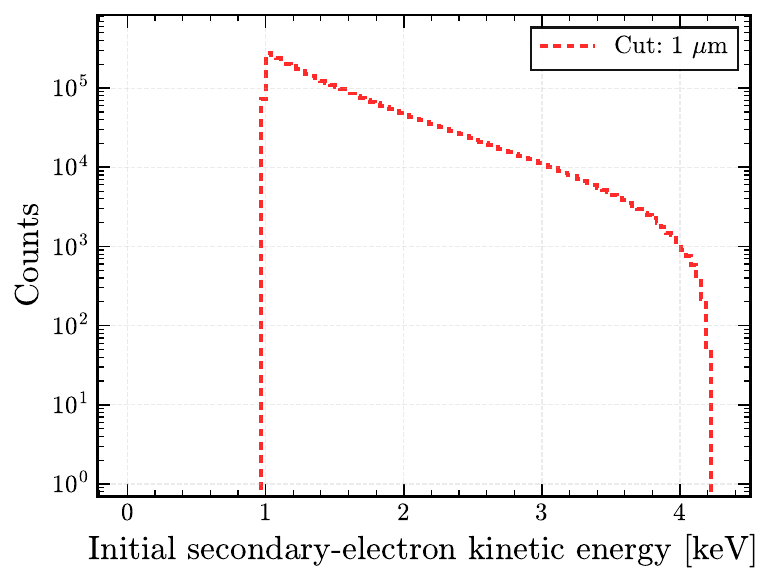}
    \end{minipage}
    \caption{
    Step-level energy-deposition diagnostics for 7.68~MeV $\alpha$ particles.
    The left panel shows the deposited energy in each primary step, and the
    right panel shows the initial kinetic energy of secondary electrons
    generated by the primary track.  No secondary-electron spectrum appears for
    the 10 and 100~$\mu$m cuts because the maximum transferable electron
    kinetic energy is below the corresponding production thresholds.
    }
    \label{fig:edep-secondary-alpha}
\end{figure}

\begin{figure}[!htb]
    \centering
    \begin{minipage}{0.49\linewidth}
        \centering
        \includegraphics[width=\linewidth]{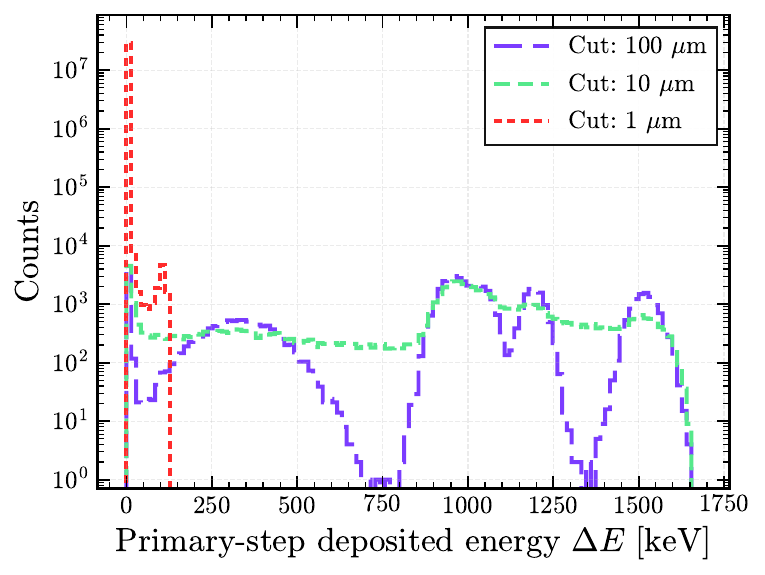}
    \end{minipage}
    \hfill
    \begin{minipage}{0.49\linewidth}
        \centering
        \includegraphics[width=\linewidth]{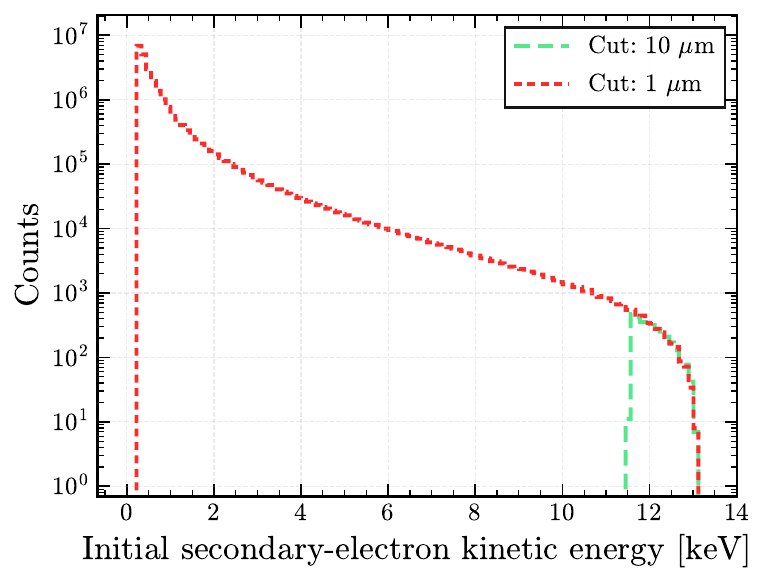}
    \end{minipage}
    \caption{
    Step-level energy-deposition diagnostics for 6~MeV protons.  At the
    100~$\mu$m production cut no primary-generated secondary electron appears
    in this sample because the maximum transferable electron kinetic energy is
    below the corresponding production threshold, while at 10~$\mu$m a small
    population of above-threshold secondary electrons is already produced.
    }
    \label{fig:edep-secondary-proton}
\end{figure}

\begin{figure}[!htb]
    \centering
    \begin{minipage}{0.49\linewidth}
        \centering
        \includegraphics[width=\linewidth]{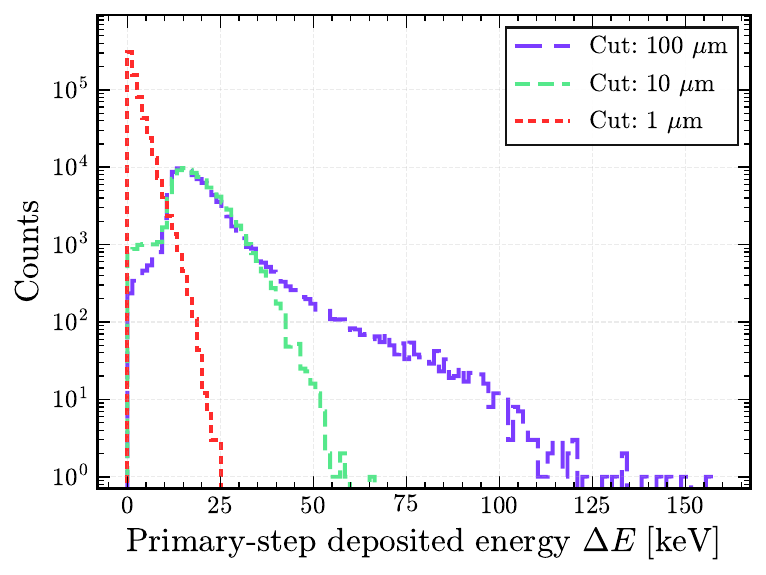}
    \end{minipage}
    \hfill
    \begin{minipage}{0.49\linewidth}
        \centering
        \includegraphics[width=\linewidth]{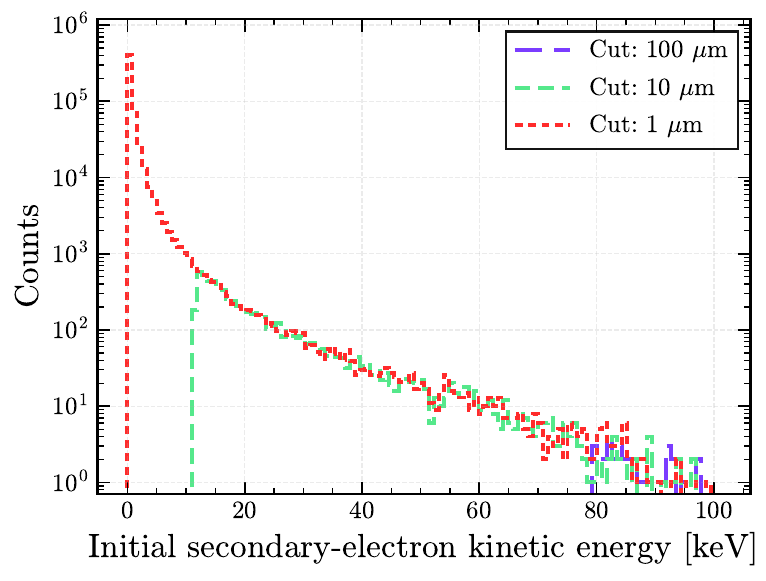}
    \end{minipage}
    \caption{
    Step-level energy-deposition diagnostics for 0.2~MeV electrons.  Secondary
    electrons are produced for all three production cuts, although the resolved
    low-energy population increases strongly when the production cut is lowered.
    }
    \label{fig:edep-secondary-electron}
\end{figure}

We next examine how the deposited energy is split between primary steps and
explicit secondary electrons.  Figures~\ref{fig:edep-secondary-alpha},
\ref{fig:edep-secondary-proton}, and \ref{fig:edep-secondary-electron} show the
primary-step deposited-energy distributions and the initial kinetic-energy
distributions of secondary electrons generated directly by the primary track.

For a heavy charged particle, the maximum kinetic energy transferable to a
delta electron is
\begin{equation}
T_{\rm max} =
\frac{2m_e c^2\beta^2\gamma^2}
{1+2\gamma m_e/M+(m_e/M)^2},
\end{equation}
where $M$, $\beta$, and $\gamma$ are the mass and relativistic parameters of
the incident particle.  Here $m_e$ is the electron mass, $M$ is the incident
heavy-particle mass, $\beta=v/c$, and
$\gamma=(1-\beta^2)^{-1/2}$.  For a 7.68~MeV $\alpha$ particle this gives
$T_{\rm max}\simeq4.2$~keV, consistent with the endpoint of the secondary
electron spectrum in Fig.~\ref{fig:edep-secondary-alpha}.  This endpoint is
below the Geant4 production thresholds corresponding to the 10 and
100~$\mu$m cuts, so explicit secondary electrons are produced only for the
1~$\mu$m cut in this $\alpha$ sample.  For 6~MeV protons, the corresponding
value is about 13~keV.  It is below the 100~$\mu$m threshold but above the
10~$\mu$m threshold, explaining why no secondary electron is generated at the
100~$\mu$m cut while a small above-threshold population is visible at
10~$\mu$m in Fig.~\ref{fig:edep-secondary-proton}.  For primary
electrons, electron-electron (Moller) scattering can transfer up to about half
of the incident kinetic energy to the secondary electron.  For the 0.2~MeV
electron sample, the corresponding upper scale is therefore about 100~keV, and
secondary electrons are visible for all three cuts in
Fig.~\ref{fig:edep-secondary-electron}.  The secondary-electron spectra also
show the approximate kinetic-energy
thresholds obtained by Geant4 when converting the range cuts in this material:
the 100, 10, and 1~$\mu$m cuts correspond to secondary-electron thresholds of
about 80, 11.5, and 0.2~keV, respectively.

\begin{figure*}[!htb]
    \centering
    \includegraphics[width=0.95\textwidth]{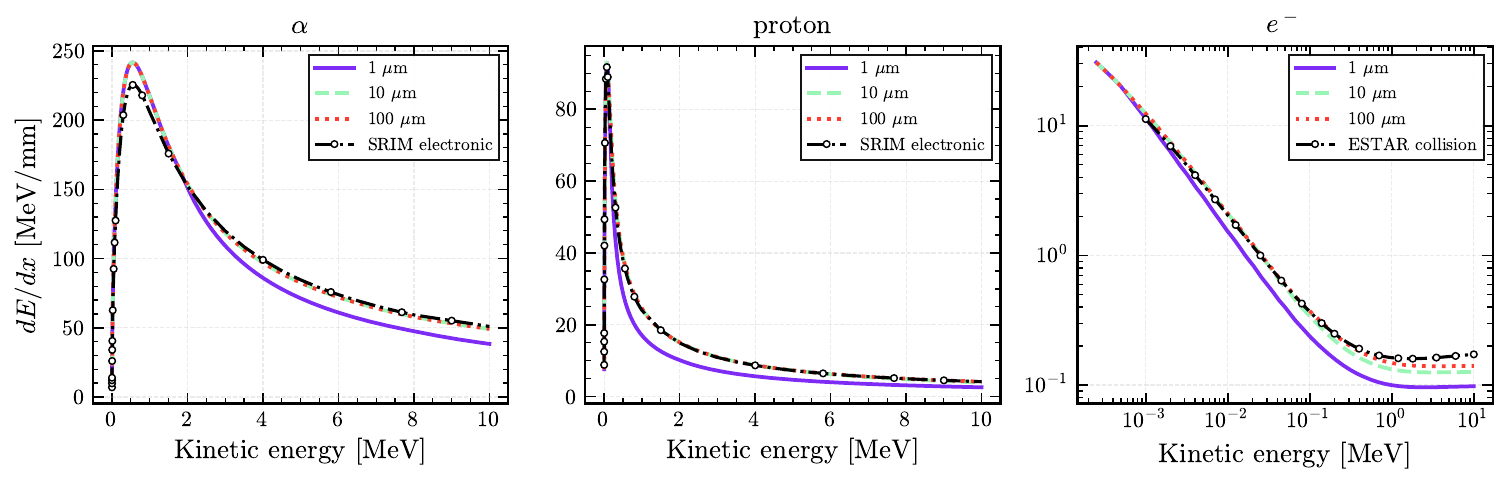}
    \caption{
    Geant4 restricted stopping-power curves for $\alpha$ particles, protons,
    and electrons in the simulated LAB-based scintillator.  Colored curves show
    the Geant4 Livermore $dE/dx$ values for electron/positron production cuts
    of 1, 10, and 100~$\mu$m.  The black curves show SRIM electronic stopping
    powers for $\alpha$ particles and protons, and the ESTAR collision stopping
    power for electrons.
    }
    \label{fig:dedx-curves-three-particles}
\end{figure*}

The production-cut dependence is also reflected in the Geant4 stopping-power
tables.  Figure~\ref{fig:dedx-curves-three-particles} compares the Geant4
restricted $dE/dx$ curves with ESTAR and SRIM references.
The ESTAR and SRIM tables provide average stopping powers for the incident
particle; the energy transferred to ionization electrons is included in the
primary-particle stopping power rather than transported as explicit
secondary-electron tracks.  In Geant4, the production cut separates ionization
energy transfer into a restricted continuous-loss part and an above-cut
secondary-production part \cite{Geant4_energy_loss}.  Schematically, for an incident particle with kinetic energy $E$, the restricted
stopping power $S_{\rm restr}(E,T_{\rm cut})$ can be written as
\begin{equation}
S_{\rm restr}(E,T_{\rm cut}) =
n \int_0^{T_{\rm cut}} T\,\frac{d\sigma(E,T)}{dT}\,dT ,
\end{equation}
where $T$ is the kinetic energy transferred to an ionization electron in a
single collision, $d\sigma(E,T)/dT$ is the corresponding differential
ionization cross section, and $n$ is the density of scattering centers in the
material.  The factor $T$ appears because this integral is an energy-loss
quantity: all sub-threshold transfers, from zero up to the production
threshold $T_{\rm cut}$, are weighted by their transferred energy and included
in the restricted stopping power of the parent track.  In contrast, the
above-cut secondary-production cross section
$\sigma_{\rm above}(E,T_{\rm cut})$ is proportional to
\begin{equation}
\sigma_{\rm above}(E,T_{\rm cut}) =
\int_{T_{\rm cut}}^{T_{\rm max}}
\frac{d\sigma(E,T)}{dT}\,dT .
\end{equation}
Here $T_{\rm max}$ is the maximum transferable energy allowed by two-body
kinematics.  This second integral is not weighted by $T$ because it represents
the probability, or cross section, for producing an explicit secondary
electron with kinetic energy above $T_{\rm cut}$.
Thus lowering the cut moves part of the energy loss from the restricted
primary-track $dE/dx$ into explicit secondary-electron tracks.  For 10 and
100~$\mu$m cuts the Geant4 curves remain close to ESTAR or SRIM for the three
representative particles, while the 1~$\mu$m cut exposes the redistribution of
energy into low-energy secondary tracks most clearly.

This redistribution matters directly for Birks' law.  Taking the $\alpha$
sample as an example, at the 100~$\mu$m cut the deposited energy in a primary
step is typically about 1-4 MeV , corresponding to primary-track stopping
powers of order $100$--$200~{\rm MeV/mm}$.  The explicitly produced
secondary electrons shown in the 1~$\mu$m cut data, however, have kinetic energies in the
1--4~keV range and electron stopping powers of order
$10~{\rm MeV/mm}$.  Treating this energy as separate electron tracks therefore
changes the local $dE/dx$ values entering Eq.~\eqref{eq:simulation-step-birks},
and can substantially change
the Birks-corrected visible energy.

\section{Extraction of the Birks' coefficient}
\label{sec:kb-extraction}

Figure~\ref{fig:track3d-alpha} shows that, when the production cut is large,
the primary charged-particle track can be represented by only a few long
Geant4 steps.  This is not ideal for extracting a Birks coefficient, because
Birks' law in Eq.~\eqref{eq:introduction-birks} is a differential empirical
relation and the step-level implementation in
Eq.~\eqref{eq:simulation-step-birks} evaluates the local quantity
$\Delta E/\Delta x$ separately for each step.  To reduce this numerical
granularity effect, the coefficient extraction below uses the samples with an
explicit maximum step length of 0.001~mm.  The production cut
is still scanned at 1, 10, and 100~$\mu$m in order to test the effect of
explicit secondary-electron production.

The experimental inputs are chosen from LAB-based scintillator measurements,
whose material compositions are close to the simulated LAB-like scintillator
summarized in Table~\ref{tab:ls-composition-comparison}.  For electrons we use
the JUNO/TUM low-energy electron quenching dataset shown in Fig.~4 of the JUNO
energy-resolution paper \cite{JUNO_energy_resolution}.  The F.~Zhang CPC
electron points are shown only as a comparison dataset.  For protons we use
the JUNO D--T-neutron elastic-scattering measurement
\cite{yangMeasurementProtonQuenching2019}.  For $\alpha$ particles we use the
SNO+ LAB1 sample.
Only the electron points below 0.2~MeV are used in the fit in order to
remove Cherenkov-light contributions to the energy response.  The
MeV-scale proton and $\alpha$ samples considered here are below the Cherenkov
threshold in the scintillator.

The three datasets are converted to dimensionless response ratios before the
fit.  The published JUNO/TUM electron quantity is already a relative response,
$E_{\rm vis}/E_{\rm true}$.  The electron experimental data are normalized at
$E_{\rm true}=0.1$~MeV \cite{JUNO_energy_resolution}. 
The JUNO proton experimental data are normalized to the 0.321~MeV electron
data \cite{yangMeasurementProtonQuenching2019}, giving
\begin{equation}
R_{p}^{\rm MC}(E;kB) =
\frac{Q_p(E;kB)/E}{Q_e(0.321~{\rm MeV};kB)/0.321~{\rm MeV}} .
\end{equation}
Here $E$ is the kinetic energy of the incident particle, $kB$ is the Birks
coefficient being fitted, and $Q_p(E;kB)$ and $Q_e(E;kB)$ are the
Birks-corrected visible energies obtained from the Geant4 step-by-step
calculation for protons and electrons, respectively.  Thus
$R_p^{\rm MC}$ is the simulated proton visible-energy fraction expressed
relative to the same low-energy electron scale used in the JUNO measurement.
The SNO+ LAB1 $\alpha$ measurement \cite{SNO+_dissertation} reports an electron-equivalent light output
after a linear electron calibration, $PH=mL+a$.  Using the LAB1 calibration
parameters $m=343.2~{\rm ch/MeV}$ and $a=-9.0~{\rm ch}$, the digitized data are
converted to
\begin{equation}
R_{\rm SNO+}^{\rm data}(E) =
A\,\frac{mL(E)+a}{m\times0.321~{\rm MeV}+a}
\frac{0.321~{\rm MeV}}{E}.
\label{eq:sno-alpha-response}
\end{equation}
In this expression, $L(E)$ is the digitized SNO+ electron-equivalent light
output at $\alpha$ kinetic energy $E$, $m$ and $a$ are the slope and intercept
of the SNO+ linear electron calibration.  Because the electron and proton
constraints are from JUNO LS whereas the $\alpha$ data are from SNO+ LAB1, the
factor $A$ aligns their absolute quenching scales by matching the SNO+ LAB1
proton response at 6~MeV to the JUNO value of about 0.5~MeV.  This gives
$A=1.024$, corresponding to a difference of only 2.4\% in the absolute
quenching scales of the two LAB-based scintillators.  This close agreement is
expected because the LS compositions used in the two experiments are nearly
identical.  The first ratio in Eq.~\ref{eq:sno-alpha-response} converts the
reported light output
to a pulse-height response relative to the 0.321~MeV electron anchor, and the
final factor converts it to a visible-energy fraction.
The simulated
$\alpha$ response is compared with
\begin{equation}
R_{\alpha}^{\rm MC}(E;kB) =
\frac{Q_{\alpha}(E;kB)}{Q_e(0.321~{\rm MeV};kB)}
\frac{0.321~{\rm MeV}}{E}.
\end{equation}
Here $Q_{\alpha}(E;kB)$ is the Birks-corrected visible energy from the
simulated $\alpha$ track.  The denominator $Q_e(0.321~{\rm MeV};kB)$ sets the
same electron calibration point as used for the proton comparison.
Using the same low-energy electron anchor for $\alpha$ particles
avoids normalizing to a high-energy electron response, where Cherenkov light
can contribute to the electron signal but not to the heavy charged-particle
response.

\begin{figure*}[!htb]
    \centering
    \includegraphics[width=1.0\textwidth]{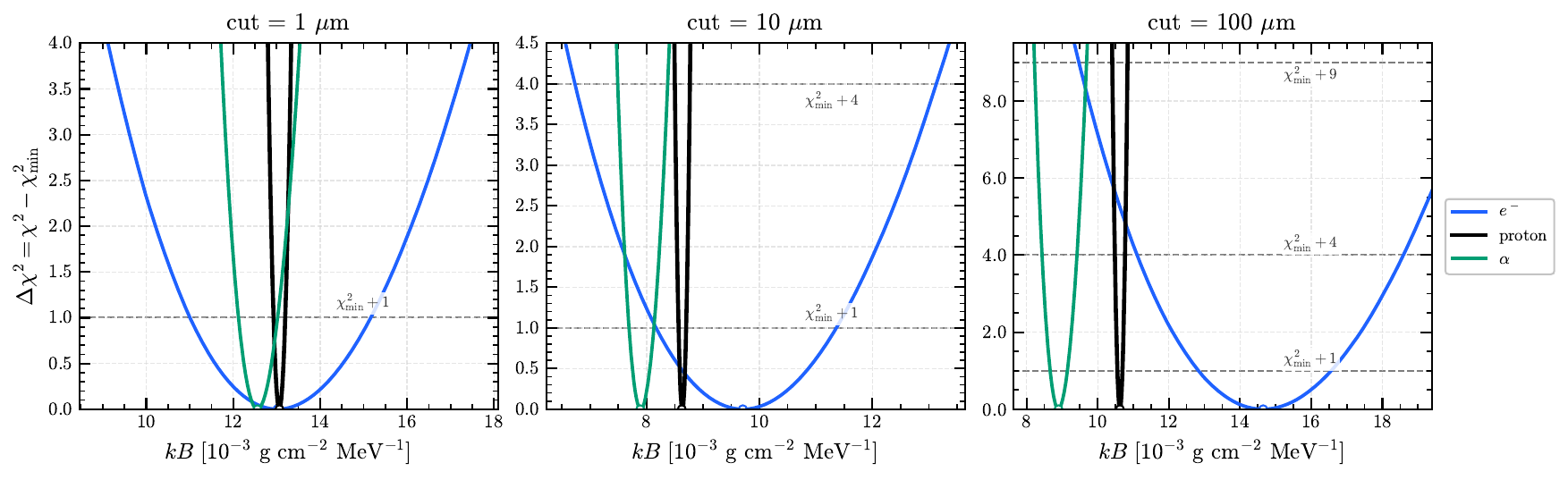}
    \caption{
    $\Delta\chi^2=\chi^2-\chi^2_{\min}$ scans for the step-limited samples.
    Each panel corresponds to one electron/positron production cut.  Horizontal dashed
    lines mark $\chi^2_{\min}+1$, $\chi^2_{\min}+4$, and
    $\chi^2_{\min}+9$, corresponding to the approximate 1, 2, and 3 standard
    deviation intervals for a one-parameter scan.
    }
    \label{fig:birks-chi2-scans-all-w}
\end{figure*}

\begin{figure*}[!htb]
    \centering
    \includegraphics[width=1.0\textwidth]{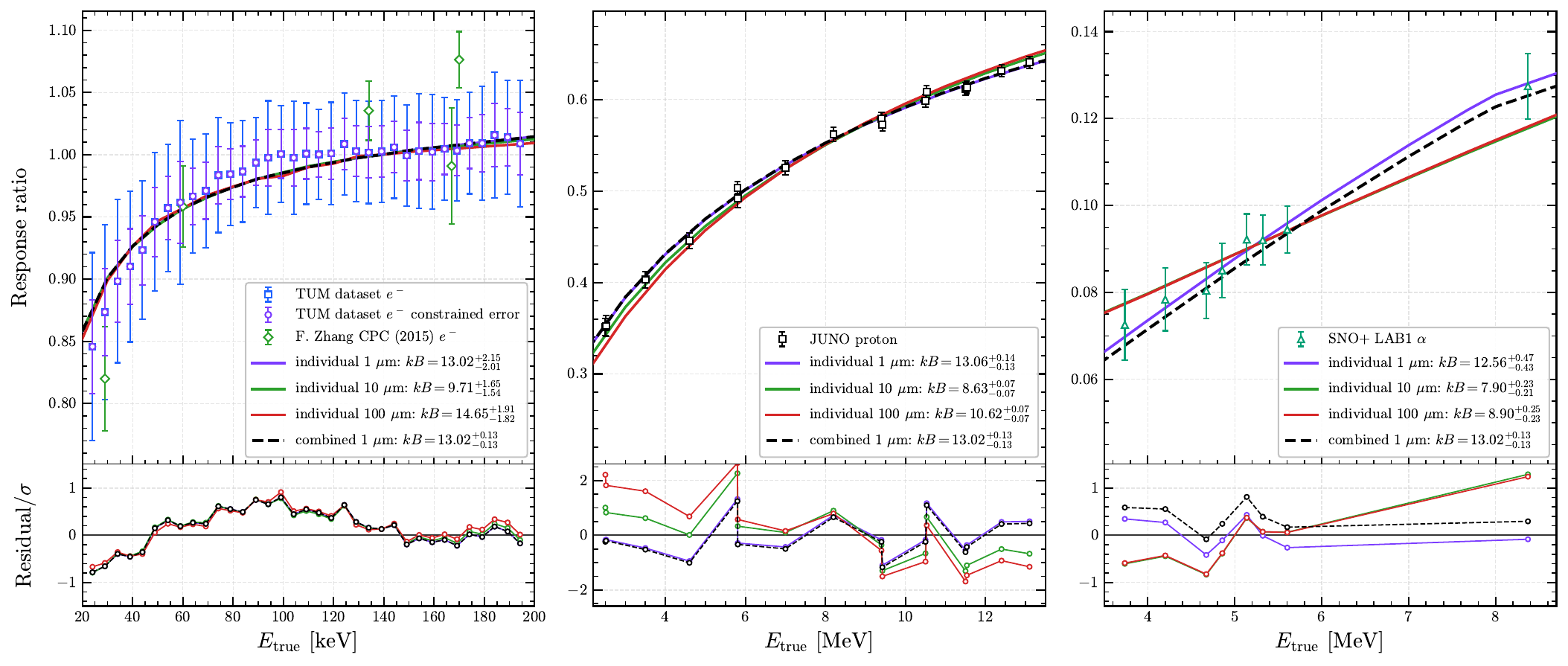}
    \caption{
    Step-limited Birks-coefficient extraction for JUNO/TUM electrons, JUNO
    protons, and SNO+ LAB1 $\alpha$ particles. The proton and $\alpha$ panels use the
    0.321~MeV electron anchor described in the text.  The colored curves show
    the individual fits at production cuts of 1, 10, and 100~$\mu$m.  The
    black dashed curve shows the combined fit with the 1~$\mu$m cut.  The
    lower panels show residual pulls for the fitted datasets,
    $(R_{\rm data}-R_{\rm MC})/\sigma$, evaluated with the uncertainties used
    in the corresponding fits.
    }
    \label{fig:birks-extraction-all-w}
\end{figure*}

\begin{table*}[!htb]
    \centering
    \scriptsize
    \renewcommand{\arraystretch}{1.15}
    \begin{tabular}{@{}lccccccccc@{}}
        \toprule
        Cut &
        \makecell{JUNO/TUM $e^-$\\publ.} &
        \makecell{JUNO/TUM $e^-$\\constr.} &
        $\chi^2/{\rm ndf}$ &
        \makecell{JUNO\\proton} &
        $\chi^2/{\rm ndf}$ &
        \makecell{SNO+ LAB1\\$\alpha$} &
        $\chi^2/{\rm ndf}$ &
        \makecell{Combined\\constr.} &
        $\chi^2/{\rm ndf}$ \\
        \midrule
        1~$\mu$m &
        $13.02^{+4.43}_{-3.90}$ &
        $13.02^{+2.15}_{-2.01}$ &
        $6.03/34$ &
        $13.06^{+0.14}_{-0.13}$ &
        $7.26/15$ &
        $12.56^{+0.47}_{-0.43}$ &
        $0.64/7$ &
        $13.02^{+0.13}_{-0.13}$ &
        $15.01/58$ \\
        10~$\mu$m &
        $9.71^{+3.43}_{-2.98}$ &
        $9.71^{+1.65}_{-1.54}$ &
        $5.87/34$ &
        $8.63^{+0.07}_{-0.07}$ &
        $14.46/15$ &
        $7.90^{+0.23}_{-0.21}$ &
        $3.22/7$ &
        $8.58^{+0.06}_{-0.06}$ &
        $32.78/58$ \\
        100~$\mu$m &
        $14.65^{+3.94}_{-3.54}$ &
        $14.65^{+1.91}_{-1.82}$ &
        $5.99/34$ &
        $10.62^{+0.07}_{-0.07}$ &
        $30.04/15$ &
        $8.90^{+0.25}_{-0.23}$ &
        $3.04/7$ &
        $10.54^{+0.07}_{-0.07}$ &
        $79.79/58$ \\
        \bottomrule
    \end{tabular}
    \caption{
    Best-fit Birks' coefficients obtained with the step-limited simulation
    sample.  Values are shown in units of
    $10^{-3}~{\rm g\,cm^{-2}\,MeV^{-1}}$.  The JUNO/TUM electron fits use the
    same central values; ``constr.'' uses the constrained error bars, while
    ``publ.'' uses the published TUM error bars.  The $\chi^2/{\rm ndf}$
    columns are shown for the constrained-error electron, proton, $\alpha$,
    and combined fits. $\chi^2/{\rm ndf}$ values are much smaller than 1
    because correlated systematic components and conservative point-by-point
    uncertainties make the effective experimental errors larger than purely
    uncorrelated statistical errors.
    }
    \label{tab:birks-fit-all-w}
\end{table*}

For each production cut and each particle species, the simulated response is
linearly interpolated both in energy and in $kB$.  The fit minimizes
\begin{equation}
\chi^2(kB) =
\sum_i \left[
\frac{R_i^{\rm data}-R_i^{\rm MC}(E_i;kB)}{\sigma_i}
\right]^2 .
\end{equation}
Here $R_i^{\rm data}$ is the normalized experimental response of the $i$th
data point, $R_i^{\rm MC}(E_i;kB)$ is the corresponding simulated response
interpolated to the measured energy $E_i$ and trial value of $kB$, and
$\sigma_i$ is the experimental uncertainty used for that point.
For the JUNO/TUM electron dataset, the published TUM uncertainties are much
larger than the scatter of the central values and give a weak constraint on
the fitted $kB$.  To test the electron shape with a more restrictive but still
conservative uncertainty, we keep the same manually adjusted central values
and define a constrained-error set by scaling the published TUM errors by 0.5,
which is comparable to the size of the F.~Zhang CPC uncertainties.  Both
electron error sets are displayed in Fig.~\ref{fig:birks-extraction-all-w};
the constrained-error set is used in the $\chi^2$ scans and in the main
combined fit.  The proton and $\alpha$ fits use the uncertainties digitized
from the corresponding experimental figures.  The individual fits use one
particle species at a time.
Table~\ref{tab:birks-fit-all-w} also lists the fits obtained with the
published TUM electron uncertainties.  In that case the individual electron
$kB$ uncertainty increases to about
$(3$--$4.5)\times10^{-3}~{\rm g\,cm^{-2}\,MeV^{-1}}$, while the common fit is
mostly constrained by the proton and $\alpha$ datasets.

Several of the fitted $\chi^2/{\rm ndf}$ values are much smaller than 1.  This
is expected because part of the quoted experimental uncertainty is correlated
among data points, and in some cases the published point-by-point errors are
conservative for the shape comparison used here.  For the JUNO/TUM electron
data, the systematic uncertainty was estimated from measurements at four
incident angles.  In the original JUNO proton and SNO+ $\alpha$ analyses,
energy-scale, beam-direction, calibration, and other systematic effects were
included through pull or nuisance parameters, while the remaining uncertainty
components were combined into the reported point-by-point errors.  As numerical
covariance matrices are not publicly available, we use these reported total
uncertainties and do not interpret the absolute $\chi^2/{\rm ndf}$ as a strict
probability; instead, the same definition is used consistently as a conservative goodness-of-fit measure and as a relative
metric for comparing production-cut settings.

The $\Delta\chi^2$ scans in Fig.~\ref{fig:birks-chi2-scans-all-w} show that
the preferred $kB$ values are strongly separated when the production cut is
large.  At the 100~$\mu$m cut, the proton and $\alpha$ simulations generate
little or no explicit low-energy secondary-electron structure, and the
electron, proton, and $\alpha$ best-fit regions are separated by more than the
$\Delta\chi^2=9$ scale.  At 10~$\mu$m the separation is reduced but still
visible.  At 1~$\mu$m, where the simulation resolves substantially more
secondary-electron tracks, the preferred values move much closer together and
a common Birks coefficient can describe the three datasets simultaneously.

The response comparison in Fig.~\ref{fig:birks-extraction-all-w} shows the
same trend in data space.  Decreasing the production cut reduces the residual
difference between simulation and measurement.  The effect is especially clear
for the $\alpha$ data: with the 10 and 100~$\mu$m cuts, the simulated curve
cannot describe the low-energy and high-energy parts of the SNO+ LAB1 dataset
with one $kB$, and the high-energy residuals reach the level of a few
standard deviations.  With the 1~$\mu$m cut, the residuals for the electron,
proton, and $\alpha$ datasets are mostly within about one standard deviation.
This indicates that a significant part of the apparent particle dependence of
the extracted Birks coefficient originates from the treatment of
secondary-electron production and step-level energy deposition in the
simulation.

\subsection{Discussion}

The Livermore low-energy electromagnetic model is used as the
configuration in this work because the JUNO detector simulation directly uses
G4EmLivermorePhysics \cite{JUNO_energy_resolution}.  As model
cross-checks, we also consider G4EmPenelopePhysics and the default
G4EmStandardPhysics constructor, which are listed in the Geant4 Physics
List Guide as the well-established electromagnetic configurations recommended for different applications  \cite{Geant4_EMPhysics}.
Apart from replacing the electromagnetic constructor, the material, geometry,
production cuts, maximum step length, experimental inputs, and fitting
procedure are unchanged.

\begin{figure}[!htb]
    \centering
    \includegraphics[width=\linewidth]{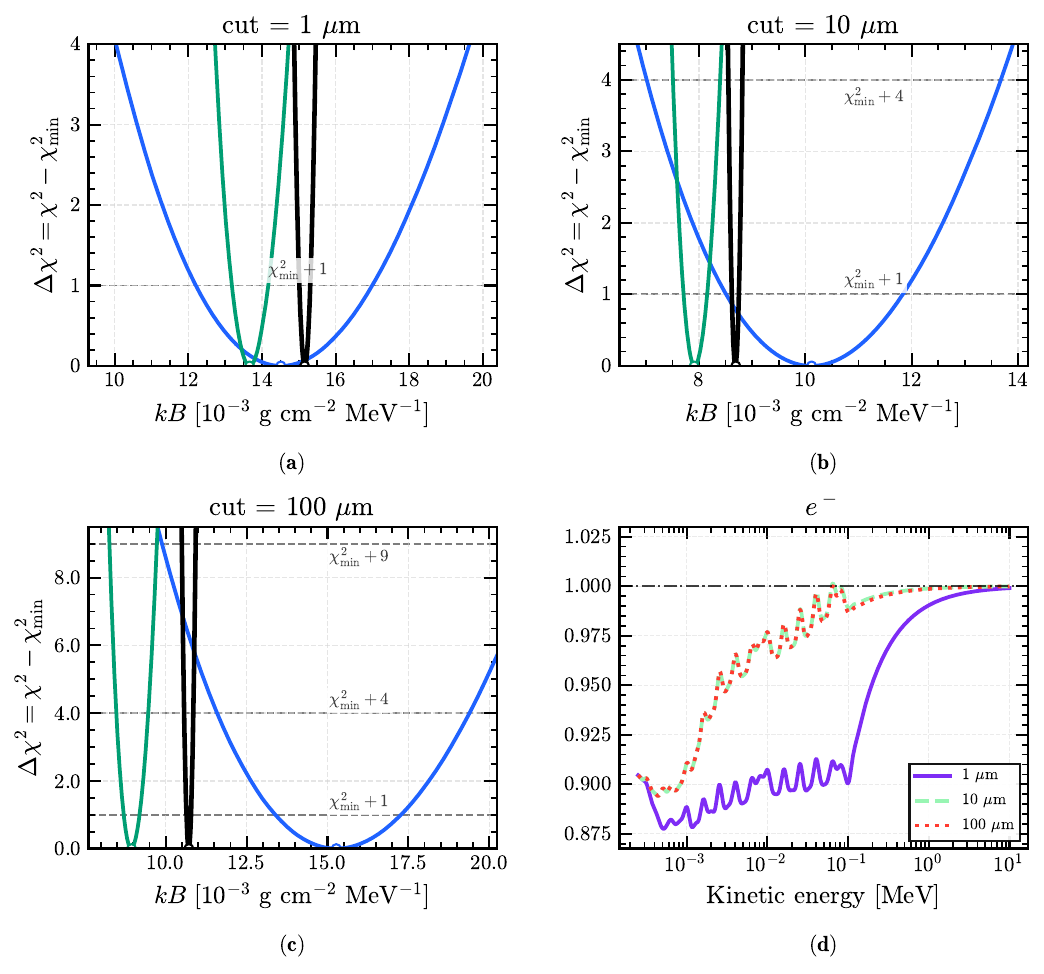}
    \caption{
    Cross-check with the Penelope model.  Panels (a)--(c) show the
    $\Delta\chi^2$ scans at electron/positron production cuts of 1, 10, and
    100~$\mu$m, respectively; the blue, black, and green solid curves denote
    the electron, proton, and $\alpha$ fits.  Panel (d) shows the ratio of the
    Penelope to Livermore restricted electron stopping powers for the three
    production cuts.
    }
    \label{fig:penelope-em-model-check}
\end{figure}

\begin{figure}[!htb]
    \centering
    \includegraphics[width=\linewidth]{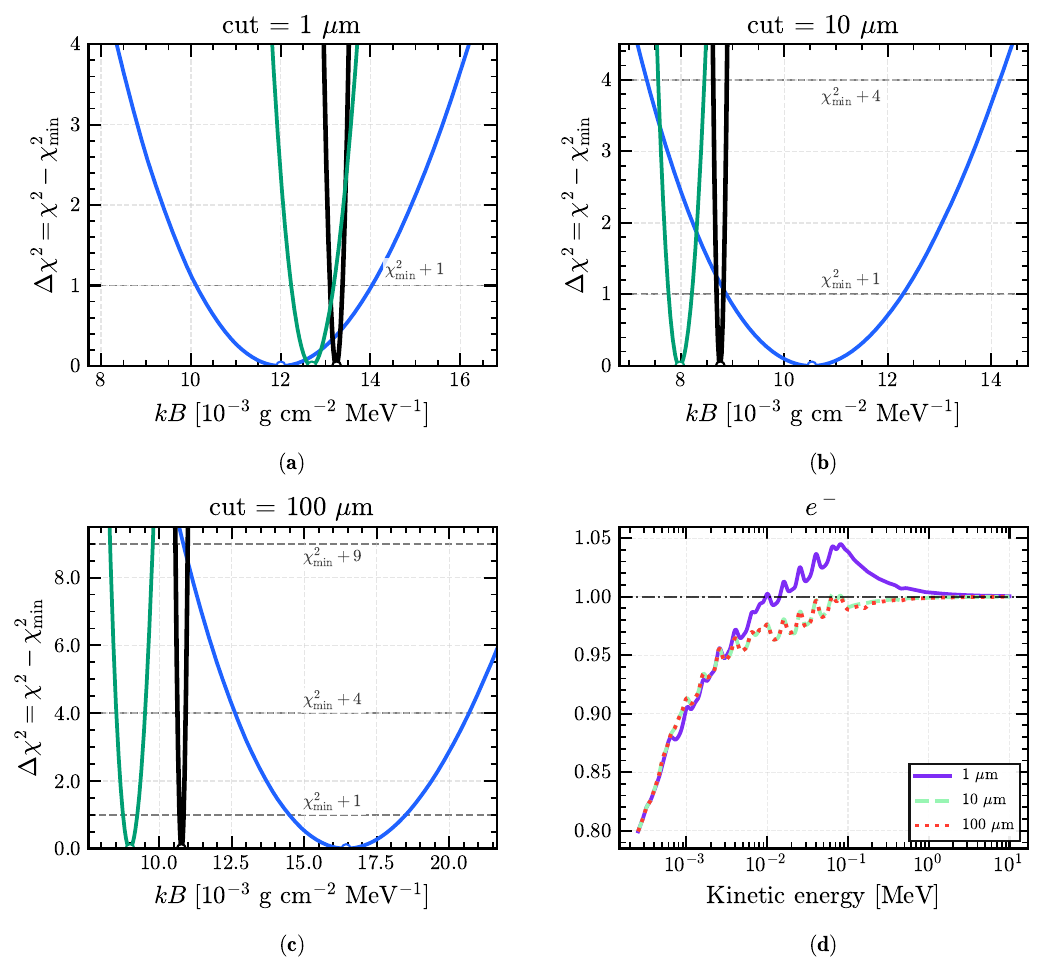}
    \caption{
    Cross-check with the StandardPhysics model.  Panels (a)--(c) show
    the $\Delta\chi^2$ scans at electron/positron production cuts of 1, 10,
    and 100~$\mu$m, respectively; the blue, black, and green solid curves
    denote the electron, proton, and $\alpha$ fits.  Panel (d) shows the ratio
    of the StandardPhysics to Livermore restricted electron stopping powers
    for the three production cuts.
    }
    \label{fig:standard-em-model-check}
\end{figure}

\begin{table*}[!htb]
    \centering
    \scriptsize
    \renewcommand{\arraystretch}{1.15}
    \begin{tabular}{@{}llccccccccc@{}}
        \toprule
        Model &
        Cut &
        \makecell{JUNO/TUM $e^-$\\publ.} &
        \makecell{JUNO/TUM $e^-$\\constr.} &
        $\chi^2/{\rm ndf}$ &
        \makecell{JUNO\\proton} &
        $\chi^2/{\rm ndf}$ &
        \makecell{SNO+ LAB1\\$\alpha$} &
        $\chi^2/{\rm ndf}$ &
        \makecell{Combined\\constr.} &
        $\chi^2/{\rm ndf}$ \\
        \midrule
        \multirow{3}{*}{Penelope} &
        1~$\mu$m &
        $14.52^{+5.12}_{-4.49}$ &
        $14.52^{+2.47}_{-2.32}$ &
        $8.95/34$ &
        $15.16^{+0.15}_{-0.15}$ &
        $8.76/15$ &
        $13.66^{+0.51}_{-0.48}$ &
        $0.70/7$ &
        $15.07^{+0.14}_{-0.14}$ &
        $25.67/58$ \\
        & 10~$\mu$m &
        $10.12^{+3.56}_{-3.10}$ &
        $10.12^{+1.72}_{-1.60}$ &
        $5.87/34$ &
        $8.69^{+0.07}_{-0.07}$ &
        $15.23/15$ &
        $7.92^{+0.23}_{-0.21}$ &
        $3.21/7$ &
        $8.63^{+0.06}_{-0.06}$ &
        $34.51/58$ \\
        & 100~$\mu$m &
        $15.27^{+4.11}_{-3.68}$ &
        $15.27^{+2.00}_{-1.89}$ &
        $6.98/34$ &
        $10.71^{+0.07}_{-0.07}$ &
        $31.50/15$ &
        $8.94^{+0.26}_{-0.23}$ &
        $3.04/7$ &
        $10.63^{+0.07}_{-0.07}$ &
        $84.79/58$ \\
        \midrule
        \multirow{3}{*}{Standard} &
        1~$\mu$m &
        $12.01^{+4.19}_{-3.66}$ &
        $12.01^{+2.03}_{-1.89}$ &
        $6.62/34$ &
        $13.24^{+0.14}_{-0.14}$ &
        $8.70/15$ &
        $12.69^{+0.50}_{-0.45}$ &
        $0.77/7$ &
        $13.19^{+0.14}_{-0.14}$ &
        $17.60/58$ \\
        & 10~$\mu$m &
        $10.52^{+3.64}_{-3.19}$ &
        $10.52^{+1.76}_{-1.65}$ &
        $6.05/34$ &
        $8.76^{+0.07}_{-0.07}$ &
        $15.54/15$ &
        $7.97^{+0.24}_{-0.21}$ &
        $3.26/7$ &
        $8.70^{+0.06}_{-0.06}$ &
        $35.84/58$ \\
        & 100~$\mu$m &
        $16.44^{+4.24}_{-3.83}$ &
        $16.44^{+2.06}_{-1.96}$ &
        $3.30/34$ &
        $10.77^{+0.07}_{-0.07}$ &
        $31.77/15$ &
        $8.98^{+0.26}_{-0.23}$ &
        $3.09/7$ &
        $10.69^{+0.07}_{-0.07}$ &
        $84.53/58$ \\
        \bottomrule
    \end{tabular}
    \caption{
    Best-fit Birks' coefficients obtained with the Penelope and StandardPhysics models.  Values are shown in units
    of $10^{-3}~{\rm g\,cm^{-2}\,MeV^{-1}}$.  The definitions of ``publ.''
    and ``constr.'', the normalization anchors, and the quoted
    $\chi^2/{\rm ndf}$ values are the same as in
    Table~\ref{tab:birks-fit-all-w}.
    }
    \label{tab:birks-fit-em-models}
\end{table*}

The Penelope and StandardPhysics results preserve the main trend of the
nominal analysis.  Decreasing the production cut increases the fraction of
low-energy secondary electrons that are explicitly tracked, reduces the
overall spread among the particle-dependent best-fit $kB$ values, and improves
the common fit from $\chi^2/{\rm ndf}=84.79/58$ at 100~$\mu$m to $25.67/58$
at 1~$\mu$m for Penelope and from $84.53/58$ to $17.60/58$ for StandardPhysics (Table~\ref{tab:birks-fit-em-models}).  The $\alpha$ fits at 1~$\mu$m
also give the smallest $\chi^2$ in both models, indicating an improved
simultaneous description of the low- and high-energy parts of the measured
$\alpha$ response.

The agreement at 1~$\mu$m is nevertheless weaker with Penelope than with the
nominal Livermore model, whereas StandardPhysics gives similar agreement.  In
Fig.~\ref{fig:penelope-em-model-check}(a), the proton and $\alpha$ minima remain
separated, and the Penelope common-fit value $25.67/58$ is larger than the
Livermore value $15.01/58$ in Table~\ref{tab:birks-fit-all-w}; the StandardPhysics value is $17.60/58$ [Fig.~\ref{fig:standard-em-model-check}(a)].  This
difference is associated with the low-energy electron stopping power: for the
1~$\mu$m cut, the Penelope and StandardPhysics restricted electron $dE/dx$ is about 10\% to 20\% lower
than the Livermore value over much of the relevant low-energy region
[Fig.~\ref{fig:penelope-em-model-check}(d), Fig.~\ref{fig:standard-em-model-check}(d)].  The different electron
stopping powers change the quenching of explicitly transported secondary
electrons and therefore shift the fitted $kB$ values in a particle-dependent
way.

Therefore, further research is still needed on different electromagnetic
models at the low-energy region.  Besides, for electrons, the experimental
uncertainties reported with the JUNO/TUM measurement are relatively large
\cite{JUNO_energy_resolution}, leading to a broad $\chi^2$ profile and a weak
constraint on $kB$ in all three electromagnetic models.  More precise low
energy electron quenching measurements, together with further validation of
the electromagnetic-model dependence, will be needed to obtain a stronger
electron constraint.

In addition, because lowering the production cut to improve the simulation precision also
increases the computational cost, we measured the runtime at different cuts.
In the Livermore transport-and-quenching simulation without optical photons,
the average times at 100, 10, and 1~$\mu$m are $1.75\times10^{-3}$,
$1.98\times10^{-3}$, and $7.64\times10^{-3}$~s/(event MeV), respectively.
Thus, the 10~$\mu$m sample is 13\% slower than the 100~$\mu$m sample, while
the 1~$\mu$m sample is 4.37 times slower.  The absolute time remains small
compared with a complete simulation, in which optical-photon generation
and transport dominate the runtime. For example, in JUNO simulation, it typically requires about
0.6~s/(event MeV) \cite{lin2023deferred}.  In exploratory 1000-event JUNOSW
tests with optical processes enabled, reducing the cut from 100 to 1~$\mu$m
changed the total runtime by factors of approximately 1.0 for 6~MeV protons
and 1.6 for 7.68~MeV $\alpha$ particles.  These results indicate that a
1~$\mu$m cut is computationally feasible for detector simulation, although
its cost should be validated for each production workflow.

\section{Summary and outlook}

We studied the extraction of Birks' coefficient for electrons, protons, and
$\alpha$ particles in LAB-based liquid scintillators using a Geant4 step-level
implementation of ionization quenching.  The comparison shows that the
production and tracking of low-energy secondary electrons is part of the
effective quenching-model definition.  When these secondary tracks are treated
more explicitly, the apparent particle dependence of the fitted Birks
coefficient is reduced, and the selected electron, proton, and $\alpha$
datasets can be described by a common effective value.  In particular, the
combined fit with the Livermore model, a 1~$\mu$m production cut, and a
1~$\mu$m maximum step length gives
$kB=(13.024^{+0.132}_{-0.128})\times10^{-3}~{\rm g\,cm^{-2}\,MeV^{-1}}$
as a common effective coefficient for the three particle species.  Moreover,
the $\alpha$ data no longer require an additional correction term beyond canonical
Birks' law: after lowering the production cut and tracking the secondary
electrons explicitly, a one-parameter Birks description can reproduce both the
low- and high-energy parts of the measured $\alpha$ response.

The same track-structure information is also relevant beyond the mean
quenching response.  Because secondary-electron production changes how the
deposited energy is divided among primary and secondary tracks, it can affect
event-by-event fluctuations of the visible energy and therefore the
non-Poisson contribution to the energy resolution.  It is also important
for particle identification.  For example, pulse-shape discrimination (PSD)
depends on the relative prompt and delayed scintillation components, which are
controlled by the local excitation density along the ionization column. A
more consistent treatment of secondary-electron transport can therefore help
future detector models describe light-yield quenching, energy-resolution, and PSD within a common track-structure framework.

\section{Acknowledgement}
The work is supported by the National Key R\&D Program of China under Grant (2023YFA1606102).

\bibliographystyle{apsrev}
\bibliography{LSQuenching}

\end{document}